\documentclass[a4paper,twocolumn,11pt,accepted=2026-03-27]{quantumarticle}
\pdfoutput=1
\usepackage[utf8]{inputenc}
\usepackage[english]{babel}
\usepackage[T1]{fontenc}
\usepackage{amsmath}
\usepackage{hyperref}
\usepackage{booktabs}
\usepackage{tikz}
\usepackage{lipsum}

\usepackage{graphicx}
\usepackage{dcolumn}
\usepackage{bm}
\usepackage{ltxtable}
\usepackage{amsmath, amsfonts, amssymb, amsthm}
\usepackage{physics}
\usepackage{graphicx}
\usepackage{bm}
\usepackage{braket}
\usepackage{mathtools}
\usepackage{algorithm,algorithmic}
\usepackage{tikz}
\usetikzlibrary{quantikz2}
\usepackage{braket}
\usepackage{caption}
\usepackage{multirow}
\usepackage{optidef}
\usepackage{soul,xcolor}
\usepackage{longtable}

\usepackage{hyperref}

\newtheorem{thm}{Theorem}

\setstcolor{purple}

\begin{document}


\title{Elevating Variational Quantum Semidefinite Programs for Polynomial Objectives}

\author{Iria Wang}
\affiliation{Department of Physics, Harvard University, Cambridge, MA 02138, USA}
\author{Robin Brown}
\affiliation{Autonomous Systems Laboratory, Stanford University, Stanford, CA 94305, USA}
\author{Taylor L. Patti}
\affiliation{NVIDIA, Santa Clara, California 95051, USA}
\author{Anima Anandkumar}
\affiliation{Department of Computing + Mathematical Sciences (CMS), California Institute of Technology (Caltech), Pasadena, CA 91125, USA}
\author{Marco Pavone}
\affiliation{Autonomous Systems Laboratory, Stanford University, Stanford, CA 94305, USA}
\affiliation{NVIDIA, Santa Clara, California 95051, USA}
\author{Susanne F. Yelin}
\affiliation{Department of Physics, Harvard University, Cambridge, MA 02138, USA}

\begin{abstract}
    Many practically important NP-hard optimization problems are inherently higher-order polynomial optimizations, which are typically addressed using approximation algorithms.
    Classical relaxations express polynomial objectives over a polynomial basis and solve the resulting quadratic objective as a semidefinite program, which can significantly inflate problem size and degrade approximation behavior.
    Variational quantum analogues to classical semidefinite programs (vQSDPs) are near-term formulations geared towards quadratic objectives.
    We introduce Product-State Lifting (PSL), a simple product-register encoding that upgrades any vQSDP with basis-state encoding to tackle $k$-degree polynomial optimization.
    This upgrade requires only a linear increase in resources with constraints constant in $k$.
    As a worked example, we pair PSL with the recently-proposed vQSDP with the Hadamard test and approximate amplitude constraints~\cite{Patti_2023}, and outline an application to Max-$k$SAT.
    PSL maintains the device-friendly structure of vQSDPs while making polynomial degree a linear resource parameter, offering a general path from quadratic to polynomial optimization without the constraint growth typical of classical relaxations.
\end{abstract}

\maketitle

\section{Introduction} \label{sec:introduction}

Recent research in quantum computation has suggested that quantum resources may enable computational advantages. 
Such an advantage could have profound implications in several fields, including drug development, materials science, machine learning, and operations research~\cite{Biamonte_Wittek_Pancotti_Rebentrost_Wiebe_Lloyd_2017, Cao_Romero_Olson_Degroote_Johnson_Kieferova_Kivlichan_Menke_Peropadre_Sawaya_etal._2019, Montanaro_2016, Preskill_2018}.
One application of interest is to the ubiquitious NP-hard combinatorial optimization problems~\cite{Korte_Vygen_2018, Berg_Hyttinen_Jarvisalo_2019, Vandenberghe_Boyd_1996, ApproximationAlgorithms}, whose complexities scale exponentially with problem size.
Assuming P $\neq$ NP and NP $\nsubseteq$ BQP, there does not exist any polynomial-time classical or quantum algorithm that exactly solves an NP-hard problem.

As an alternative solution, more efficient yet less accurate techniques known as \emph{approximation algorithms} are typically used.
Improving the performance of these algorithms remains a central research area in classical optimization theory~\cite{Korte_Vygen_2018,Cai_Zhang_2020, maxsat_2016, Parrilo_2003, Goemans_Williamson_1995, Berg_Hyttinen_Jarvisalo_2019, HickeyB22, Princeton_SOS, Prajna_Papachristodoulou_Parrilo_2002, Vandenberghe_Boyd_1996, Harrach_2022, Gepp_Harris_Vanstone_2020}.
Classical semidefinite programming (SDP) is a classical approximation algorithm targeting quadratic optimization with linear constraints. 
SDPs are often used for NP-hard problems that are naturally expressible as quadratic optimizations over integer-valued variables, e.g. MaxCut~\cite{Goemans_Williamson_1995}.

Significant work has been done to develop a quantum analogue to classical SDPs in pursuit of a speed-up.
Quantum SDPs (QSDPs)~\cite{Brandao_Kalev_Li_Lin_Svore_Wu_2019, van_Apeldoorn_Gilyen_2019, Apeldoorn_Gilyen_Gribling_Wolf_2020, Brandao_Svore_2017, Brandao_Kueng_Franca_2022} are favored for their theoretical guarantees but are unsuitable for noisy near-term devices.
Variational approaches are specifically tailored for near-term devices~\cite{Ebadi_Keesling_Cain_Wang_Levine_Bluvstein_Semeghini_Omran_Liu_2022, Albash_Lidar_2018, Farhi_Goldstone_Gutmann_Sipser_2000, Gibney_2017, Kadowaki_Nishimori_1998, Farhi_Goldstone_Gutmann_2014, Arrazola_Bergholm_Bradler_Bromley_Collins_Dhand_Fumagalli_Gerrits_Goussev_Helt_etal._2021, PhysRevA.101.010301, PhysRevResearch.6.033257}, and recently proposed variational quantum semidefinite programs (vQSDP) preserve theoretical guarantees, including bounds on the SDP true optimum~\cite{Chen_Westerheim_Holmes_Luo_Nuradha_Patel_Rethinasamy_Wang_Wilde_2025,Patel_Coles_Wilde_2021,Patti_2023}.
These vQSDPs are formulated by rewriting all constraints as penalty terms in the problem's objective function, and variationally minimizing its value.

While SDPs are a well-studied cornerstone for quadratic optimization, many problems are inherently higher-order, arising naturally as $k$-degree polynomial objectives.
Examples of higher-order polynomial optimizations with linear constraints include MaxSAT and its variants, the polynomial knapsack problem, and $k$-local Ising models with fixed magnetization.
Solving these problems by reducing them to quadratic optimizations, for instance by using the well-known classical sum-of-squares (SOS) approach~\cite{Parrilo_2003, Prajna_Papachristodoulou_Parrilo_2002}, can significantly inflate problem size and degrade approximation behavior. These disadvantages motivate methods that treat polynomial optimization natively, rather than forcing it through a quadratic proxy.

In this work, we present a simple multi-register, product-state construction that ``lifts'' any basis-state vQSDP from quadratic to $k$-degree with resource growth linear in $k$.
This approach treats polynomial objectives natively.
Odd-degree terms may be made even through the introduction of a dummy variable, and degree-2$d$ terms are encoded using $d$ identical $n$-qubit registers. 
Crucially, our tensor-product formulation inherently enforces \emph{algebraic consistency}, resulting in a constraint burden that is unchanged in $k$.
We call this method \emph{Product State Lifting (PSL)}.

In this manuscript, we apply PSL to the recently proposed vQSDP with Hadamard tests and approximate amplitude constraints~\cite{Patti_2023}.
The new algorithm is applied to the Max-$k$SAT problem,
which is a prototypical degree-$k$ polynomial objective over discrete variables and a standard benchmark family for higher-order optimization. This choice is primarily illustrative, and the PSL mechanism applies more broadly to polynomial encodings.

Simulations of our approach are run for Max-3SAT instances with up to 110 variables and 1500 clauses, with results demonstrating that our approach beats classical SOS for small problems, is tractable for large problems, and is competitive with classical heuristics.
Our reported performance is obtained via classical simulation of the variational loop, suggesting that the PSL lifting itself already yields a competitive variational template. Implementing the same objective-evaluation primitive on quantum hardware would remove the need to explicitly represent $2^n$-dimensional state vectors and could enable scaling to regimes where classical simulation becomes prohibitive.

\section{Preliminaries} \label{sec:preliminaries}

In this section, we review relevant existing approximation techniques. 
This includes two standard classical relaxation techniques---semidefinite programming (SDP) and sum-of-squares (SOS) programming---which serve as both algorithmic foundations and practical benchmarks for this work. 
By comparison, heuristic solvers (often based on local search or variational algorithms) aim to find feasible solutions to optimization problems quickly without guarantees on solution quality \cite{Cai_Zhang_2020, maxsat_2016, HickeyB22}. 

\subsection{Classical Relaxations} \label{sec:classical_relaxations}

Relaxations replace a hard discrete or non-convex problem with a tractable convex surrogate whose optimum certifies a bound on the true objective.
The relaxation strictly broadens the solution space, upper bounding the \emph{feasible region} of the original problem.
The solution corresponding to the upper bound may be outside the feasible region, so a rounding step then maps the surrogate solution back to a feasible solution space.
This yields an approximate, but possibly sub-optimal, solution and lower bound.
Semidefinite programming (SDP) relaxations are standard techniques for quadratic objectives, whereas sum-of-squares (SOS) moment hierarchies are standard for polynomials.

\subsubsection{Semidefinite Programming} \label{sec:semidefinite_programming}

SDP is a class of convex optimization problems that aims to extremize a linear objective function over symmetric positive semidefinite matrices $\mathbb{S}^+$ under a set of constraints. 
With diverse applications in control theory and combinatorial optimization~\cite{Parrilo_2003, Vandenberghe_Boyd_1996, Harrach_2022, Gepp_Harris_Vanstone_2020, Lemon_So_Ye_2016}, SDPs are often used to bound solutions to NP-hard maximization problems.

In general, the relaxed SDP form of these problems is given as
\begin{mini}
    {X \in \mathbb{S}^+}{\expval{W,X} \hspace{2.5cm}}
    {}{\label{eqn:classical_SDP}}
    \addConstraint{\expval{A_\mu, X} = b_\mu \, \forall \mu \leq M,}
\end{mini}
where $W$ is an $N \times N$ symmetric matrix encoding a specific problem instance, matrix $A_\mu$ and scalar $b_\mu$ describe problem constraints, and the angled brackets denote the trace inner product
\begin{align}
    \expval{A, B} = \Tr(A^T B) = \sum_{i,j = 1}^N A_{i,j} B_{i,j}
\end{align}
for some matrices $A$ and $B$. These SDPs are exactly solvable using classical techniques such as interior point methods \cite{Vandenberghe_Boyd_1996}.

\subsubsection{Sum-of-Squares} \label{sec:sum_of_squares}

SOS programming is a core discipline in operations research \cite{Princeton_SOS}.
Here, we provide a cursory overview of the relevant SOS formulae, providing a more detailed treatment in the Appendix.

SOS programming determines whether a degree-$2D$ polynomial function $F(y)$, where $y = \{y_i\}_{i \in \{ 0, 1, ..., N-1\}}$, can be factored as a sum of squares. 
We express $F(y)$ as 
\begin{align}
    F(y) = b^T Q b,
\end{align}
where $Q$ is a symmetric matrix and 
\begin{align}
    b^T = \begin{pmatrix} 1 & y_1 & y_2 & ... & y_n & y_1y_2 &  ... &y_n^D \end{pmatrix} \label{eqn:basis_vector}
\end{align}
is a \emph{polynomial basis}. 
This basis enables the generalization to higher-degree polynomials. 
The matrix $Q$ can be interpreted as a weight matrix, where $F(y) = b^T Q b$ is a sum of squares if and only if $Q$ is positive semidefinite. SOS is therefore solvable by SDP \cite{Princeton_SOS}. 

To address polynomial optimization, we define $F(y) \equiv - p(y) - \gamma$ where $p(y)$ is a polynomial objective and $\gamma$ is a real number. Writing $F(y)$ as SOS is an algebraic certificate of nonnegativity, implying $F(y) = - p(y) - \gamma >0$. Then,  $- \gamma$ is an upper bound on $p(y)$.

\vspace{5mm}

In the SOS hierarchy, higher degree is handled by expanding the polynomial basis and adding constraints to keep moments consistent.
For instance, the product of the polynomial basis entries representing $y_i$ and $y_j$ is not necessarily equal to the entry representing $y_iy_j$, unless otherwise enforced.
Thus, SOS does not inherently preserve \emph{algebraic consistency}, which must be addressed with additional constraints or rounding procedures~\cite{vanMaaren_vanNorden_Heule_2008}.
While SOS is a standard and useful theoretical tool, increasing the degree of the polynomial causes the constraints and the polynomial basis to inflate unfavorably, such that even moderately-sized problems quickly become intractable in practice.

\subsection{Polynomial Max-$k$SAT Formulations} \label{sec:polynomial_maxksat_formulations}

Maximum Satisfiability (MaxSAT) is a quintessential example of an NP-hard combinatorial optimization problem.
MaxSAT can be used in practical settings to solve optimization problems arising in data analysis and machine learning, and studying this problem provides insight into complexity theory and approximation algorithm performance~\cite{Berg_Hyttinen_Jarvisalo_2019, Cai_Zhang_2020}.

In MaxSAT, we are given a list of boolean (true or false) clauses, each consisting of boolean variables.
The goal is to find a truth assignment of these variables that maximizes the number of satisfied clauses, i.e., clauses that evaluate to True.
MaxSAT instances are typically written, without loss of generality, in conjunctive normal form (CNF), where each clause consists of boolean variables or their negations connected by the logical ``or'' (denoted $\lor$) and each clause is connected by the logical ``and''.
Any propositional formula, or equation with some truth value, may be written in CNF form \cite{Howson1997}.

Max-$k$SAT is a special case of MaxSAT, where each clause is restricted to contain $k$ variables. Max-$k$SAT is NP-hard, simpler to analyze, generalizable to MaxSAT, and well-studied \cite{Berg_Hyttinen_Jarvisalo_2019, Cai_Zhang_2020, Goemans_Williamson_1995, maxsat_2016, HickeyB22}, making it an attractive benchmark for quantum computing approaches to NP-hard problems.
In general, Max-$k$SAT can be formulated as a degree-$k$ polynomial optimization problem over integer-valued variables.

\subsubsection{Max-2SAT} \label{sec:max2sat}

We consider a Max-2SAT instance on $N - 1$ boolean variables, which are represented by $\{x_i\}_{i \in \{1, ..., N-1\}}$\footnote{We note that $N - 1$ is a deliberate choice to simplify notation when we introduce an auxiliary variable.}.
A Max-2SAT instance is a series of clauses written in CNF.
Each clause is written as $(x_i \lor x_j)$, where either or both of $x_i$ and $x_j$ may be replaced by their negations (denoted $\neg x_i$ and $\neg x_j$, respectively). 

We define a set of $N$ variables $\{y_i\}_{i \in \{0, 1, ..., N-1\}}$ where $y_i \in \{1, -1\}$ for all $i \in \{0, 1, ..., N-1\}$. 
The $y_0$ defines the truth value, such that $x_i$ is true if $y_i = y_0$ and false if $y_i \neq y_0$. 

We write $v(C) = 1$ if a clause $C$ is true and $v(C) = 0$ if $C$ is false. 
Then,
\begin{align}
    v(x_i) = \frac{1 + y_0 y_i}{2} \text{, } v(\neg x_i) = \frac{1 - y_0 y_i}{2},
\end{align}
and
\begin{align}
\begin{split}
    v(x_i \lor x_j) & = 1 - v(\neg x_i) v(\neg x_j) \\
    & = \frac{1 + y_0 y_i}{4} + \frac{1 + y_0 y_j}{4} + \frac{1 - y_i y_j}{4}.
    \label{eqn:max2sat_v}
\end{split}
\end{align}
The value of other clauses are similarly expressed. 
If $x_i$ or $x_j$ are negated, then we replace the corresponding $y_i$ or $y_j$ with $-y_i$ or $-y_j$.

The maximum number of satisfiable clauses is written as a maximization over $\sum_\ell v(C_\ell)$ where $C_\ell$ is the $\ell$th clause in the given list of clauses. 
This is a combination of terms in Eq.~\eqref{eqn:max2sat_v},
\begin{maxi}
    {}{\frac{3}{4}n_C - \sum_{i<j} y_i y_j W_{i,j}\hspace{1.5cm}}
    {}{\label{eqn:max2sat_quadratic_form}}
    \addConstraint{y_i \in \{ - 1, 1 \}, \forall i \in \{0,...,N-1\}},
\end{maxi}
where $n_C$ is the number of clauses, $W_{i,j}$ are constants defined by the specific problem instance, and $i,j \in \{0, 1, ..., N-1\}$. 
While $y_0$ has a separate definition from other variables in $\{y_i\}_{i \in \{1, ..., N-1\}}$, we can treat $y_0$ the same as $\{y_i\}_{i \in \{1, ..., N-1\}}$ once this quadratic function is determined.

Eq.~\eqref{eqn:max2sat_quadratic_form} is a quadratic optimization problem. 
Other NP-hard problems such as MaxCut and MaxBisection may be similarly expressed as quadratic optimization problems. 
The conversion from the NP-hard problem instance to a quadratic optimization problem is done in polynomial time \cite{Goemans_Williamson_1995, ApproximationAlgorithms}.

The objective consists of only constant and quadratic (i.e., even-degree) terms. 
If we were to eliminate $y_0$ (for example, having $y_i = 1$ indicate $x_i$ as true and $y_i = -1$ indicate $x_i$ as false for $i \in \{1, ..., N-1\}$), then the resulting polynomial objective would contain linear terms.
Even-degree terms are favorable as they allow variable matrices to be defined as an outer product between vectors of variables. 
The role of $y_0$ is to formulate the objective with only even-degree terms. 
In general, any polynomial can be made even by introducing a variable analogous to $y_0$.

Eq.~\eqref{eqn:max2sat_quadratic_form} can be solved as an SDP. The variables $y_i$ are relaxed to take on a vector form $v_i \in S_{N}$, where $S_{N} \subset \mathbb{R}^{N}$ is the unit sphere in $N$ dimensions.
Then, Eq.~\eqref{eqn:max2sat_quadratic_form} becomes
\begin{mini}
    {X \in \mathbb{S}^+}{\expval{W, X} \hspace{1cm}}
    {}{\label{eqn:max2sat_sdp}}
    \addConstraint{X_{i,i} = 1,  \quad \forall i \leq N.}
\end{mini}
where $W$ is a matrix with elements $W_{i,j}$ and $X$ is a matrix with elements $X_{i,j} = v_i^T v_j$.

To obtain an approximate solution within the feasible space of the Max-2SAT instance, unrounded solutions $v_i \in S_N$ are rounded to $y_i \in \{1, -1\}$ by determining the side of a random hyperplane they fall on. 

This relaxation method is called the Goemans-Williamson (GW) approximation algorithm for Max-2SAT \cite{Goemans_Williamson_1995}.
It is proven that the GW algorithm will deliver solutions with an expected value of at least 0.87856 times the optimal number of satisfied clauses. 

\subsubsection{Max-3SAT} \label{sec:max3sat}

In Max-3SAT, clauses are written in the form $(x_i \lor x_j \lor x_k)$, where any of $x_i$, $x_j$, and $x_k$ may be replaced by their negations. 
By incorporating $y_0$ analogously to Max-2SAT, the truth value of the clause $(x_i \lor x_j \lor x_k)$ is expressed as
\begin{align}
    \begin{split}
        &v(x_i \lor x_j \lor x_k) = 1 - v(\neg x_i)v(\neg x_j)v(\neg x_k) \\
        & = \frac{1 + y_0 y_i}{8} + \frac{1 + y_0 y_j}{8} + \frac{1 + y_0 y_k}{8} + \frac{1 - y_i y_j}{8} \\
        & \hspace{5mm} + \frac{1 - y_i y_k}{8} + \frac{1 - y_j y_k}{8} + \frac{1 + y_0 y_i y_j y_k}{8}.
    \end{split}
    \label{eqn:max3sat_v}
\end{align}
The value of other clauses are similarly expressed. 
If any of $x_i$, $x_j$, and $x_k$ are negated, then we replace the corresponding $y_i$, $y_j$, or $y_k$ with $-y_i$, $-y_j$, or $-y_k$. 

The maximum number of satisfiable clauses is written as a maximization over $\sum_\ell v(C_\ell)$,
\begin{maxi}
    {}{\frac{7}{8}n_C - \sum_{i<j} y_i y_j W_{i,j}^{(1)}- \sum_{i<j<q<l} y_i y_j y_q y_l W_{ij,ql}^{(2)}}
    {}{\label{eqn:max3sat_quadratic_form}}
    \addConstraint{y_i \in \{ - 1, 1 \}, \forall i \in \{0,...,N-1\}},
\end{maxi}
where $W_{i,j}^{(1)}$ and $W_{ij,ql}^{(2)}$ are constants defined by the specific problem instance. 
The result is a non-convex degree-4 polynomial objective, consisting of only even-degree terms.

We note that we may obtain a degree-3 polynomial if we eliminate $y_0$. Classical SOS can target degree-3 polynomials, but formulating polynomials with even-degree terms using $y_0$ is more suitable to our quantum approach.

\subsubsection{Max-kSAT} \label{sec:maxksat}

Max-$k$SAT may be naturally formulated as a degree-$k$ polynomial without $y_0$, and as a degree-2$D$ polynomial with $y_0$ where $D \equiv \text{ceil}(k/2)$. 
The maximum degree of the polynomial objective is $2D$, and we also define $d \in \{1, ..., D\}$ such that $2d$ denotes the degree of individual terms in the objective. 
For example, in Max-3SAT, $D = 2$. In the degree-4 polynomial objective, we have constant, degree-2, and degree-4 terms, such that $2d = 0$, $2$, and $4$ respectively. Because constant terms do not impact the solution of an optimization problem, they are dropped and $d \in \{1, 2\}$. 

Since any polynomial can be made to have only even-degree terms using $y_0$ and our proposed algorithm makes use of the even-degree characteristic, we focus on degree-2$D$ polynomials in our discussions.

In this manuscript, Max-$k$SAT is used as a representative polynomial optimization objective since it is a standard testbed for relaxation-based methods. 
Our goal is not to propose a Max-$k$SAT-specific heuristic, but to present PSL as a general primitive to elevate vQSDP with basis-state encodings for polynomial objectives.

\subsection{Quantum Semidefinite Programs} \label{sec:quantum_semidefinite_programs}

Quantum SDP solvers promise powerful guarantees, such as polynomial-time approximation under oracle models.
However, QSDPs require deep circuits, coherent block-encodings, and sometimes QRAM-like access, rendering them intractable for noisy near-term devices~\cite{Brandao_Kalev_Li_Lin_Svore_Wu_2019, van_Apeldoorn_Gilyen_2019, Apeldoorn_Gilyen_Gribling_Wolf_2020, Brandao_Svore_2017, Brandao_Kueng_Franca_2022}.
Variational quantum approaches use shallow circuits and a hybrid structure that is catered toward near-term devices.
Recently, variational quantum SDPs (vQSDPs) that maintain theoretical guarantees, including bounds on the SDP true optimum, have been proposed~\cite{Chen_Westerheim_Holmes_Luo_Nuradha_Patel_Rethinasamy_Wang_Wilde_2025,Patel_Coles_Wilde_2021,Patti_2023}.
These vQSDPs are formulated by rewriting all constraints as penalty terms in the problem's objective function, and variationally minimizing its value.

As a focused study, we will highlight the key characteristics of the recent vQSDP with the Hadamard test and approximate amplitude constraints~\cite{Patti_2023} applied to Max-2SAT.

\section{Our Approach to Max-2SAT} \label{sec:our_approach_to_maxksat}

As a variational quantum analogue to SDPs,~\cite{Patti_2023} finds heuristic solutions for problems with $N \leq 2^n$ variables using only $O(n)$ qubits, $O(1)$ quantum measurements, and $O(\textrm{poly}(n))$ classical calculations.
The loss is determined using observables on the quantum state $\ket{\psi} = U_V \ket{0}^{\otimes n}$, and is minimized via the variational quantum circuit $U_V$.

We remark that, as with other variational quantum algorithms, minimizing loss over $U_V$ is a nonconvex variational optimization and may suffer from trainability issues at large $n$.
These issues include barren plateaus, which are effects of concentration of measure that occur with randomly initialized parameters and circuits forming Haar random 2-designs or higher.
While we do not claim iteration-complexity guarantees, several mitigation strategies have fortunately already been proposed.
Standard barren plateau mitigation strategies include entanglement regularization~\cite{Patti_Najafi_Gao_Yelin_2021} and selective initialization~\cite{Liu2023Mitigating}.
Barren plateaus also only become a possibility with very large problems, as $n$ grows logarithmically with problem size.

In this section, we extend the approach of~\cite{Patti_2023} to Max-2SAT, including some slight simplifications of the original formulation that do not alter the overall approach, solution quality, or results of ~\cite{Patti_2023}.
Beginning with the SDP formulation in Eq.~\eqref{eqn:max2sat_sdp}, we replace the matrix $X$ with the density matrix $\rho = \ket{\psi}\bra{\psi}$, where $\ket{\psi} = \begin{pmatrix} \psi_0 & \psi_1 & ... & \psi_{2^n-1} \end{pmatrix}^T$ in the computational basis.
In particular, $\psi_i$ corresponds to $y_i$ for all $i \in \{0, 1, ..., N-1\}$, where we recall $N \leq 2^n$.
The elements $\rho_{i,j}$ of $\rho$ represent $y_i y_j$, such that the quantum equivalent of Eq.~\eqref{eqn:max2sat_sdp} is\footnote{To ensure that the dimensions of $W$ and $\rho$ match, $W$ may be ``padded'' with $2^n - N$ rows and $2^n - N$ columns of zeroes. In this work, the notation is the same for padded and non-padded versions of a matrix.}
\begin{mini}
    {\rho} {\expval{W, \rho} \hspace{1.5cm}} {} {\label{eqn:max2sat_qsdp}} \addConstraint{\rho_{i,i} = 2^{-n}, \quad \forall i \leq N.}
\end{mini}
The objective function may be written as an expectation value $\expval{W, \rho} = \Tr(W \rho) = \bra{\psi} W \ket{\psi}$, and $W$ may be used to generate a unitary $U_{W} = e^{i \alpha W} = 1 + i \alpha W - O(\alpha^2)$.
Then, the value of the objective function $\bra{\psi} W \ket{\psi}$ may be approximated by using a Hadamard test to evaluate $\Im(\bra{\psi} U_{W} \ket{\psi}) \approx \alpha \bra{\psi} W \ket{\psi}$.

The Hadamard test is a quantum computing subroutine that estimates $\Im(\bra{\psi} U_{W} \ket{\psi})$ and other expectation values in the loss function.
To generalize this subroutine we could simply replace $\ket{\psi}$ with an arbitrary $n$-qubit state and $U_{W}$ with an arbitrary $n$-qubit unitary. 
The circuit used to execute the Hadamard test is shown in Figure~\ref{fig:hadamard_test}, which shows that the Hadamard test uses only a single expectation value on an ancilla qubit $\expval{\sigma_{\text{anc.}}^z}_{W} \equiv \Im(\bra{\psi} U_{W} \ket{\psi})$
instead of the $N \leq 2^n$ expectation values that would be otherwise required to fully characterize $\rho$. 

We note that the variational circuit $U_V$ is designed such that the elements $\psi_i$ of $\ket{\psi}$ must be real. 
Thus, the objective $\bra{\Psi} W \ket{\Psi}$ is real, and $\Im(\bra{\psi} U_{W} \ket{\psi}) \approx \alpha \bra{\psi} W \ket{\psi}$.
Further discussion on this approximation is in the original manuscript \cite{Patti_2023}.

\begin{figure*}
    \centering
    \includegraphics[width=.8\textwidth]{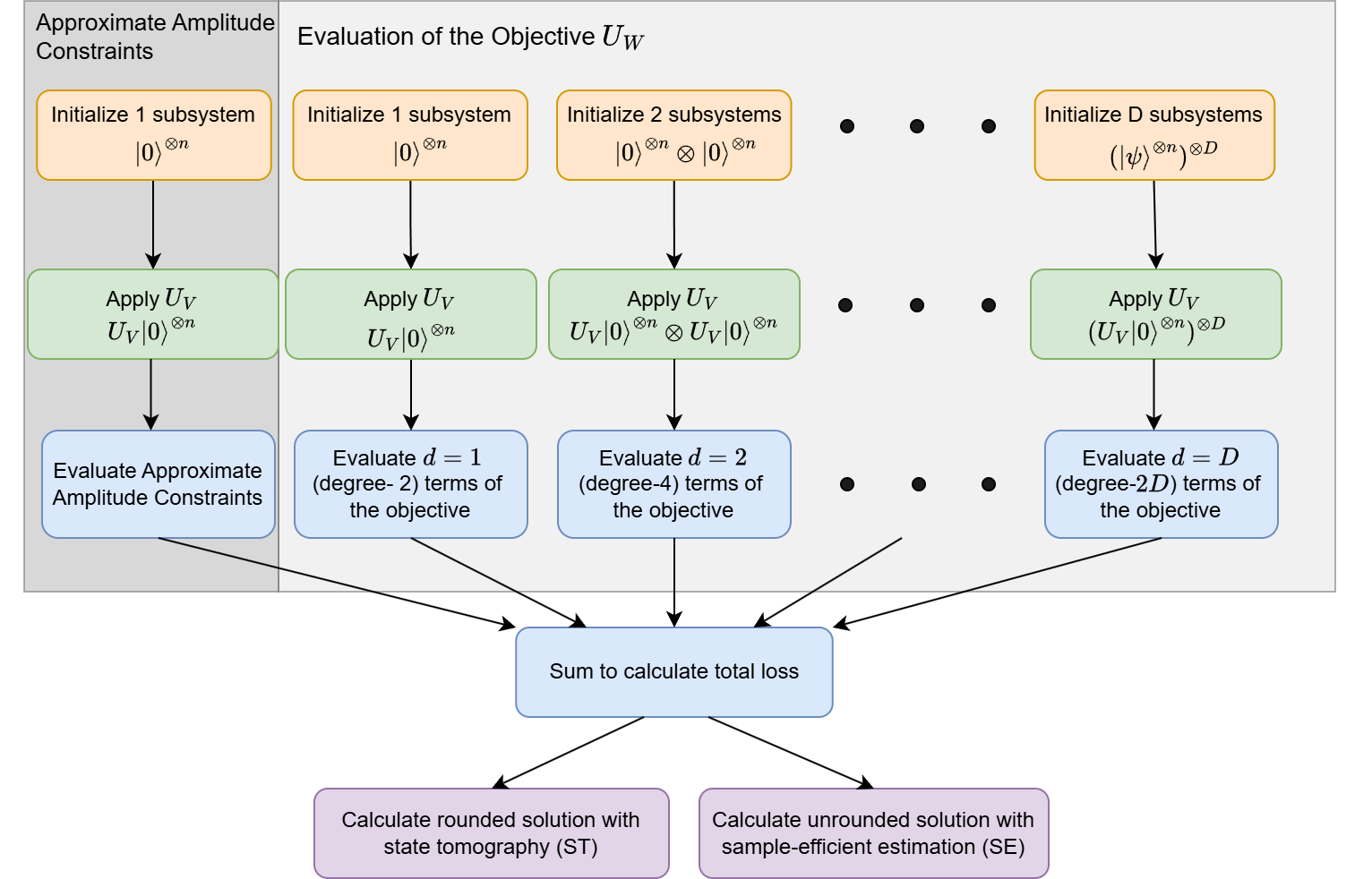}
    \captionsetup{justification=raggedright,singlelinecheck=false}
    \caption{\textbf{Our proposed approach.} Our approach is a variational algorithm, such that an $n$-qubit variational circuit $U_V$ is applied to quantum subsystems (green) and iteratively modified to minimize a specified loss function. The loss approximates a degree-2$D$ polynomial optimization problem with up to $N \leq 2^n$ variables using expectation values calculated from the total quantum system (blue). The solution is extracted by conducting full state tomography (ST) and rounding the resulting quantum state, or by using sample-efficient estimation (SE) to directly estimate the unrounded solution (purple).}
    \label{fig:HTAAC-QSOS}
\end{figure*}

\vspace{5mm}

For Max-2SAT, we wish to approximately enforce the constraint $\rho_{i,i} = 2^{-n}$ from Eq.~\eqref{eqn:max2sat_qsdp}. Enforcing $\rho_{i,i} = 2^{-n}$ is equivalent to enforcing $\abs{\psi_0}^2 = \abs{\psi_1}^2 = ... = \abs{\psi_{2^n-1}}^2$ up to normalization. This yields $2^n - 1$ independent constraints, defined by $2^n - 1$ unique equations\footnote{A unique set of equations is defined here as a set in which no equation can be deduced from other equations in the set. For example, $\{a = b, b = c\}$ are unique, but $\{a=b,b=c,a=c\}$ are not unique.}. 
We can rewrite this set of $2^n - 1$ equations using the following Pauli strings of length 1 to $n$:
\begin{align}
\begin{split}
    & \expval{\sigma_a^z, \rho} = 0, \, \forall a \leq n, \, \text{yielding $\binom{n}{1}$ equations} \\
    & \expval{\sigma_a^z \sigma_b^z, \rho} = 0, \, \forall a < b \leq n, \, \text{yielding $\binom{n}{2}$ equations} \\
    &  ... \\
    & \expval{\sigma_1^z \sigma_2^z ... \sigma_n^z, \rho}, \, \text{yielding $\binom{n}{n}$ equations.}
    \label{eqn:max2sat_all_strings}
\end{split}
\end{align}
The total number of equations is 
\begin{align}
    \sum_{p = 1}^n \binom{n}{p} = \sum_{p = 0}^n \binom{n}{p} - 1 = 2^n - 1.
    \label{eqn:Pauli_expansion}
\end{align}
Each of the Pauli strings in Eq.~\eqref{eqn:max2sat_all_strings} yields an equation $\sum_i c_i \abs{\psi_i}^2 = 0$, for $c_i \in \{1, -1\}$.

From Eq.~\eqref{eqn:Pauli_expansion}, exactly enforcing the constraint requires $O(2^n)$ observables. 
To avoid this, Patti et al. \cite{Patti_2023} propose \emph{approximate amplitude constraints}, which truncate Eq.~\eqref{eqn:Pauli_expansion} at $p = 2$ such that we only consider the $O(n^2)$ Pauli strings up to length 2. 
This series may be truncated at higher $p$ to make the approximation more precise, in exchange for less-favorable polynomial resource scaling. 
The effect of this truncation is further explored in the original paper~\cite{Patti_2023}.

Since this method is approximate,~\cite{Patti_2023} includes a possible additional component to supplement the above Pauli string constraints. We introduce a diagonal matrix $P$ and a \emph{population-balancing} unitary
\begin{align}
    \begin{split}
        & U_P = e^{i \beta P} \\
        & \text{where } P_{i,i} = -\left(P_{\text{max}} - \sum_j \abs{W_{i,j}} \right), \\
        & \hspace{1cm} P_{\text{max}} = \max_i \sum_j W_{i,j}.
    \end{split}
\end{align}
This helps to offset any asymmetric weights, and the expectation value of this unitary may be estimated using a Hadamard test\footnote{We note that the population-balancing unitary serves as a supplemental constraint-regularization term~\cite{Patti_2023}. In the simulations reported here, we omitted this term from the optimized loss.}.

\vspace{5mm}

The objective function, the population-balancing unitary, and the Pauli strings may be combined as a single loss function to be minimized:
\begin{align}
    \mathcal{L} = \expval{\sigma^z_{\text{anc.}}}_{W} + \lambda \left(\sum_{\expval{\sigma, \rho} \in \mathcal{P}} \expval{\sigma, \rho} + \expval{\sigma^z_{\text{anc.}}}_{P}\right).
\end{align}
The first term $\expval{\sigma^z_{\text{anc.}}}_{W}$ results from the Hadamard test used to estimate the objective function, the second term $\sum_{\expval{\sigma, \rho} \in \mathcal{P}} \expval{\sigma, \rho}$ represents the set of Pauli strings $\mathcal{P}$, and the third term $\expval{\sigma^z_{\text{anc.}}}_{P}$ is from the Hadamard test used to estimate the population-balancing unitary. 
The second and third terms make up our approximate amplitude constraints, and are weighted by a Lagrange multiplier $\lambda$.

\vspace{5mm}

SE may be used to directly estimate a solution from $\ket{\psi}$ as
\begin{align}
    \frac{3}{4}n_C + \bra{\psi} W \ket{\psi} = \frac{3}{4}n_C + \frac{2^{n-1}}{\alpha} \expval{\sigma_{\text{anc.}}^z}_{W}.
\end{align}
The first term is constant, and the second term represents the quadratic terms in the objective evaluated with a Hadamard test.
While this solution is unrounded, it is a good approximation of a rounded solution in the limit of well-behaved constraints. 
When the constraints are exactly enforced, the solution approaches the optimal rounded solution.

Alternatively, by using ST and a simple rounding procedure, we may extract a guaranteed feasible solution. 
We assign $y_i = \text{sign}(\psi_i)$ and substitute the resulting $\{y_i\}_{i \in \{0, 1, ..., N-1\}}$ into the objective from Eq.~\eqref{eqn:max2sat_quadratic_form}. 
While this is sample-intensive on a quantum machine, its complexity does not depend on the degree of the polynomial and the result mimics classical rounded solutions.
It may also be suitable as a quantum-inspired algorithm \cite{Hakemi_Houshmand_KheirKhah_Hosseini_2024}.

\section{Our Proposed Product-State Lift} \label{sec:our_proposed_product_state_lift}

We now introduce our proposed product-state lift (PSL), which elevates any basis-state encoding to represent polynomial objectives.
In basis-state encodings, function variables are represented in the computational basis of an $n$-qubit register, such that each computational basis state corresponds to a variable as is done in~\cite{Patti_2023}.
Nothing about this variable map changes when we move to higher-degree objectives.

Any polynomial may be converted to an even-degree polynomial with only even-degree terms by introducing a dummy variable such as $y_0$, so it is only necessary to consider even-degree monomials.
Let $k$ be the degree of the polynomial without the dummy variable, and $2D = 2 \, \text{ceil}(k/2)$ be the degree of the polynomial with the dummy variable.
PSL represents each even-degree component $2d$ for $d = 1, ..., D$ by preparing $d$ identical copies of the same $n$-qubit state $\rho$.

Algebraically, entries of the product state $\rho^{\otimes d}$ coincide with the target monomials.
As an example, for $d = 2$, $(\rho \otimes \rho)_{ij,ql} = \rho_{i,j} \rho_{q,l}$ aligns with $y_i y_j y_q y_l$.
Thus, degree-$2d$ terms are encoded by the tensor product $\rho^{\otimes d}$ while the base representation of $\rho$ remains unchanged.

\vspace{5mm}

Unlike SOS, PSL naturally preserves \emph{algebraic consistency} without imposing additional constraints or procedures.
Since every register is an identical copy of $\rho$, products like $y_i y_j$ and $y_i y_j y_q y_l$ are represented by consistent products of the same base amplitudes, so the higher-order moment relations hold by construction.
Consequently, moving from quadratic to degree $k$ does not require adding new equality constraints beyond those already present at the base layer.
The product-state representation preserves these constraints instead of re-imposing them with extra variables.

Furthermore, since all $d$ registers are copies of the same $\rho$, extracting solutions with state tomography (ST) only requires the reconstruction of a single $n$-qubit system.
In other words, ST is constant in $k$.
The tomography target does not grow with degree because PSL never introduces a distinct state for higher-order terms.

\section{Our Approach to Max-3SAT} \label{sec:our_approach_to_max3sat}

As a continuation of the previous discussion on Max-2SAT, we apply our approach to Max-3SAT, simulating the algorithm and comparing it against classical benchmarks. 

Using basis-state encoding and PSL, Eq.~\eqref{eqn:max3sat_quadratic_form} becomes
\begin{mini}
    {\rho}{\overbrace{\expval{W^{(1)}, \rho}}^{\text{degree-2 terms}} + \overbrace{\expval{W^{(2)}, \rho \otimes \rho}}^{\text{degree-4 terms}} \hspace{1cm}}
    {}{\label{eqn:max3sat_qsdp}}
    \addConstraint{\rho_{i,i} = 2^{-n},  \quad \forall i \leq N}
\end{mini}
where the objective function is the sum of two expectation values, 
\begin{align}
    \begin{split}
        &\expval{W^{(1)}, \rho} + \expval{W^{(2)}, \rho \otimes \rho} \\
        & = \bra{\psi} W^{(1)}\ket{\psi} + \big(\bra{\psi} \otimes \bra{\psi}\big) W^{(2)} \big(\ket{\psi} \otimes \ket{\psi}\big).
        \label{eqn:max3sat_obj_temp}
    \end{split}
\end{align}

\begin{figure}[h]
    \centering
    \includegraphics[width=0.48\textwidth]{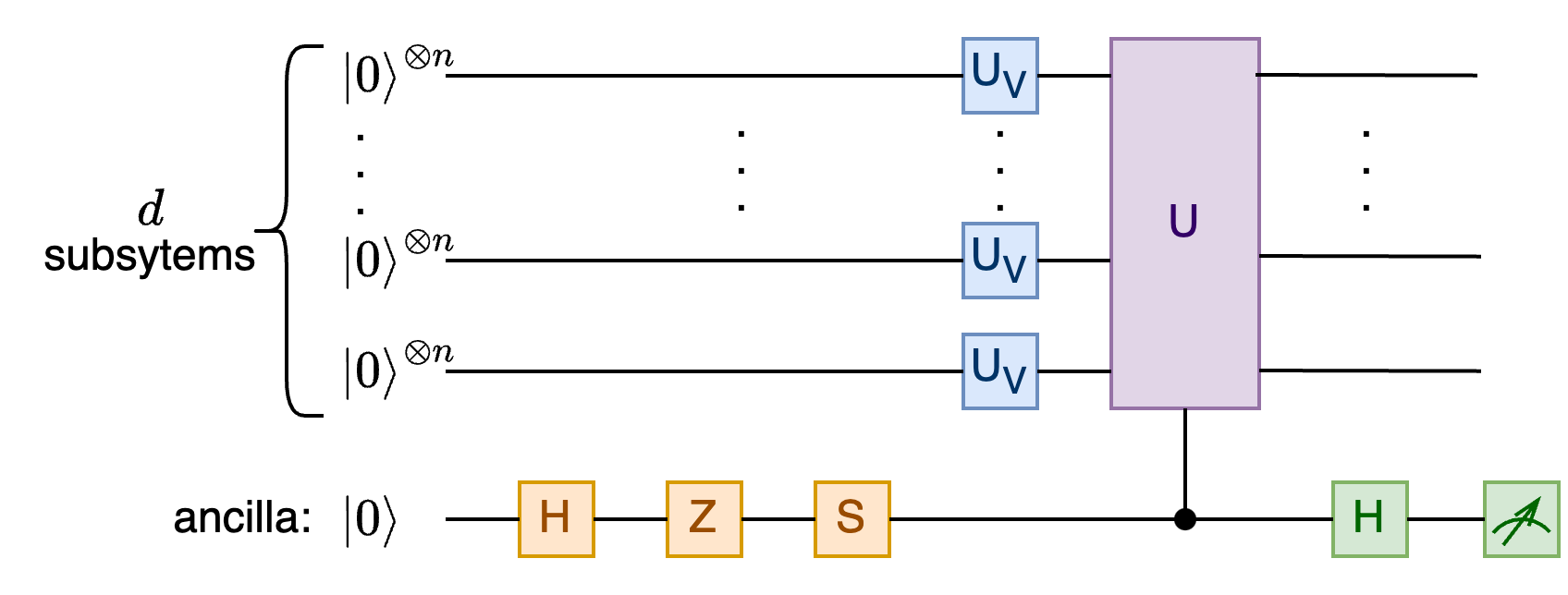}
    \captionsetup{justification=raggedright,singlelinecheck=false}
    \caption{\textbf{Hadamard test on $nd$ qubits.} The value $\Im\left(\bra{\psi}^{\otimes d} U \ket{\psi}^{\otimes d}\right)$, where $\ket{\psi} \equiv U_V \ket{0}^{\otimes n}$, may be estimated by measuring a single ancilla qubit. Specifically, $\expval{\sigma_{\text{anc.}}^z} = \Im\left(\bra{\psi}^{\otimes d} U \ket{\psi}^{\otimes d}\right)$, where $\sigma_{\text{anc.}}^z$ is the Pauli-$Z$ on the ancilla qubit.}
    \label{fig:hadamard_test}
\end{figure}

Degree-2 terms are evaluated with a Hadamard test on a single register $\rho$ and a unitary $U_{W^{(1)}} = e^{i \alpha W^{(1)}}$, similarly to Max-2SAT, 
\begin{align}
    \expval{\sigma_{\text{anc.}}^z}_{W^{(1)}} = \Im\left(\bra{\psi} U_{W^{(1)}} \ket{\psi}\right).
\end{align}
To evaluate degree-4 terms, two identical copies of the state $\rho$ are prepared on two registers. A Hadamard test is implemented by operating on both registers with a larger $2n$-qubit unitary i.e., $U_{W^{(2)}} = e^{i \alpha W^{(2)}}$,
\begin{align}
    \expval{\sigma_{\text{anc.}}^z}_{W^{(2)}} \equiv \Im\left((\bra{\psi} \otimes \bra{\psi}) U_{W^{(2)}} (\ket{\psi} \otimes \ket{\psi})\right).
\end{align}
The circuit diagram for these two Hadamard tests is shown in Figure~\ref{fig:hadamard_test}, where $d = 1$ and $d = 2$ respectively.

We want to enforce the constraint $\rho_{i,i} = 2^{-n}$ for all $i \leq N$. 
Since we encode the objective function using product states, we do not require any additional constraints {besides those required by Max-2SAT}. 
{The Pauli string constraints are identical, and the population-balancing unitary may be constructed with $W^{(1)}$ instead of $W$.}
The purpose of the population-balancing unitary is to approximate the frequency at which a variable appears, and the larger matrix $W^{(2)}$ is therefore not required.

Extending the SE method for Max-2SAT, we evaluate the unrounded {as
\begin{align}
    \frac{7}{8}n_C - \frac{2^{n-1}}{\alpha}\expval{\sigma_{\text{anc.}}^z}_{W^{(1)}} - \frac{2^{2n-1}}{\alpha}\expval{\sigma_{\text{anc.}}^z}_{W^{(2)}}
\end{align}
where the second and third terms are the results of Hadamard tests on one copy of $\rho$ for $U_{W^{(1)}} = e^{i \alpha W^{(1)}}$ and two copies of $\rho$ for $U_{W^{(2)}} = e^{i \alpha W^{(2)}}$, respectively.}

To obtain the rounded solution we require ST, and follow the same procedure outlined {as for Max-2SAT}.

\subsection{Simulations}

We test our proposed approach by simulating it for Max-3SAT on a classical system and benchmarking results against classical SOS and heuristic methods\footnote{Code can be found here: \url{https://github.com/IriaWang/QSDPs_for_Polynomial_objectives}.}.

\vspace{5mm}

To test the performance of {our approach} against SOS, we generate several Max-3SAT instances by drawing variables from a uniform distribution. SOS techniques~\cite{Prajna_Papachristodoulou_Parrilo_2002} are used to obtain classical solutions. We choose problems with 20 variables and 80, 100, 120, 140, 160, or 180 clauses because they may be approximated by SOS within reasonable time.

For each problem size, five instances are randomly generated.
For every problem instance, SOS is used to determine an upper bound.
We remark that SOS often yields very tight upper bounds~\cite{vanMaaren_vanNorden_Heule_2008, ApproximationAlgorithms}. 
However, since these upper bounds do not necessarily correspond to feasible solutions, we use a {randomized} rounding procedure described in Appendix~\ref{app:rounding_procedures} to return the SDP solutions to the feasible region.
The rounding procedure is run five times for each instance, and the best result is selected.

For each problem instance, our approach is also run 20 times with random instantiations.
Implementation details are given in the Appendix.
Both the unrounded SE and rounded ST solutions are extracted for each of the 20 runs.
The average unrounded result, best rounded solution, and average rounded solutions are found.
For comparison, we also determine the ratio between rounded solutions from our algorithm and rounded SOS solutions.

Results are provided in Table~\ref{tab:sos}. 
These solutions are averaged over all five instances of the same size, and the results of each individual instance may be found in the Appendix.

\begin{table*}
    \centering
    \begin{tabular}{|c||c|c||c||c|c|c|c|}
        \hline
        \multicolumn{1}{|c||}{Probl} & \multicolumn{2}{c||}{Classical SOS} & \multicolumn{1}{c||}{Sim Unround} & \multicolumn{4}{c|}{Sim Round}\\
        \hline \hline
        \# clauses & Unround & Round & Avg Soln & Best Soln & Best Ratio & Avg Soln & Avg Ratio \\
        \hline
        80 & 79.8 & 79.8 &  83.1 &   79.4 &  0.995 &   77.7 &  0.974 \\
        100 & 99.6 & 92.6 & 101.7 &   98.8 &  1.071 &   96.2 &  1.043 \\
        120 &      118.0 &   112.8 & 121.0 &  117.8 &  1.054 &  115.0 &  1.029 \\
        140 &      137.8 &   127.2 & 141.1 &  137.0 &  1.082 &  133.8 &  1.057 \\
        160 & 156.4 &   153.0 &    160.0 &  155.2 &  1.016 &  152.1 &  0.996 \\
        180 & 173.8 &   167.6 &    178.9 &  173.6 &  1.037 &  170.3 &  1.017 \\
        \hline 
    \end{tabular}
    \captionsetup{justification=raggedright,singlelinecheck=false}
    \caption{\textbf{Observed performance with respect to classical SOS}. Simulations of {our approach} are used to approximate solutions for randomly-generated Max-3SAT instances with 20 variables and a given number of clauses. {Results are averaged over five instances of the same size, and our approach is run 20 times for each instance with different instantiations. {Ratios are given with respect to the rounded classical SOS solutions.}}
    \label{tab:sos}}
\end{table*}

The rounded solutions from our approach are consistently close to or greater than the rounded SOS solutions, despite encoding only a rough approximation of the problem.

\vspace{5mm}

Rounding with our approach does not degrade solution quality in the way it does with SOS.
PSL naturally preserves algebraic consistency, such that the element representing $y_i$ and the element representing $y_j$ multiply to equal the element representing $y_i y_j$.
This property is naturally preserved by the product states used in our approach, but not by the rounding procedure in classical SOS. This results in a much simpler rounding scheme for our approach, and one that yields heuristics that are closer to the optimal solution, as demonstrated by our simulation result. 

Moreover, in SDP relaxations, the rank of the solution may serve as an indicator of solution quality. Many polynomial optimization problems yield a natural SDP relaxation. If the SDP relaxation has a rank-1 solution, then the relaxation is exact, and there is no gap between the upper bound and the rounded solution. Intuitively, this suggests that rank-1 solutions are more desirable \cite{Lemon_So_Ye_2016}. SDP relaxations often yield full-rank solutions. The solution to our approach is always a pure state and therefore is rank-1, suggesting that our approach may result in solutions that yield a smaller optimality gap when rounded.

\vspace{5mm}

To gauge the performance of our approach for larger problems, we draw instances with 70 to 110 variables from the 2016 MaxSAT evaluation\footnote{While more recent MaxSAT evaluation results are available, we choose the 2016 competition because it is the final year where Max-3SAT test instances are provided.}~\cite{maxsat_2016}. 
Since classical SOS is intractable for larger problems, the best known solutions are given by the winning heuristic solutions from the MaxSAT evaluation.
These solvers primarily leverage local search algorithms and problem-specific heuristics. 
Like our approach, these solvers prioritize speed over approximation ratio guarantees.
The observed performance is therefore calculated with respect to these solutions, and averaged over instances of the same size given by the MaxSAT evaluation. 

\begin{figure}
    \centering
    \includegraphics[width=.48\textwidth]{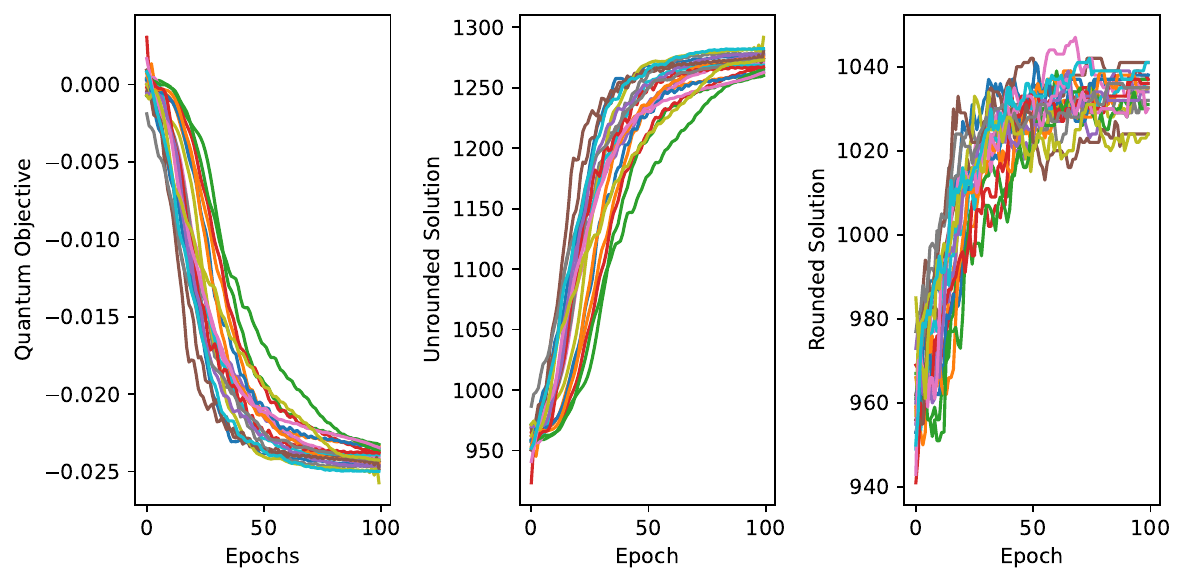}
    \captionsetup{justification=raggedright,singlelinecheck=false}
    \caption{\textbf{Loss and Observed Performance of {Our Approach} over Epochs.} Our approach is used to find a heuristic solution for a Max-3SAT instance with 110 variables and 1100 clauses. These plots show the quantum objective (left), unrounded solution (middle), and rounded solution (right) over 20 runs with random initializations.}
    \label{fig:loss_plots_big}
\end{figure}

For each given problem size, five problem instances are used.
For each instance, our approach is run 20 times with random initializations.
The average unrounded solutions, best rounded solutions, and average rounded solutions are found.
The ratios between rounded solutions from our approach and classical heuristic solutions are provided.
Results are given in Table~\ref{tab:heuristic_sim} averaged over the five problem instances of the same size.
Figure~\ref{fig:loss_plots_big} shows loss and observed performance over 100 training epochs for a Max-3SAT instance with 110 variables and 1100 clauses, {showing} that the unrounded objectives, rounded objectives, and loss are strongly correlated.

\begin{table*}
    \centering
    \begin{tabular}{|c|c||c||c||c|c|c|c|}
        \hline
        \multicolumn{2}{|c||}{Problem} & \multicolumn{1}{c||}{Classical} & \multicolumn{1}{c||}{Sim Unround} & \multicolumn{4}{c|}{Sim Round}\\
        \hline \hline
        \# var. & \# clauses & Heuristic & Avg Soln & Best Soln & Best Ratio & Avg Soln & Avg Ratio \\
        \hline 
        \multirow{8}{*}{70} & 700 & 677.4 & 815.1 & 669.8 & 0.989 &663.5 & 0.979 \\
        & 800 & 769.2 & 898.9 & 759.0 & 0.987 & 752.7 & 0.979 \\
        & 900 & 861.0 & 1000.2 & 852.6 & 0.990 & 845.1 & 0.981 \\
        & 1000 & 962.8 & 1114.7 & 940.6 & 0.977 & 935.1 & 0.971 \\
        & 1100 & 1046.2 & 1214.4 & 1039.2 & 0.993 & 1031.1 & 0.986 \\
        & 1200 & 1134.8 & 1317.9 & 1124.0 & 0.990 & 1116.8 & 0.984 \\
        & 1300 & 1228.6 & 1417.5 & 1215.2 & 0.989 & 1208.5 & 0.984 \\
        & 1400 & 1320.6 & 1521.5 & 1308.0 & 0.990 & 1298.9 & 0.984 \\
        & 1500 & 1410.0 & 1614.1 & 1397.4 & 0.991 & 1389.2 & 0.985 \\
        \hline
        \multirow{7}{*}{90} & 700 & 683.6 & 826.0 & 671.8 & 0.983 & 665.5 & 0.973 \\
        & 800 & 777.6 & 946.0 & 765.0 & 0.984 & 759.2 & 0.976 \\
        & 900 & 871.8 & 1037.5 & 858.0 & 0.984 & 850.7 & 0.976 \\
        & 1000 & 962.8 & 1146.6 & 949.0 & 0.986 & 941.0 & 0.977 \\
        & 1100 & 1055.6 & 1240.6 & 1041.4 & 0.987 & 1032.7 & 0.978 \\
        & 1200 & 1147.0 & 1353.9 & 1134.6 & 0.989 & 1124.1 & 0.980 \\
        & 1300 & 1238.2 & 1438.6 & 1220.8 & 0.986 & 1213.4 & 0.980 \\
        \hline
        \multirow{5}{*}{110} & 700 & 689.2 & 844.0 & 674.2 & 0.978 & 668.7 & 0.970 \\
        & 800 & 783.2 & 957.1 & 769.2 & 0.982 & 761.4 & 0.972 \\
        & 900 & 879.2 & 1072.2 & 863.8 & 0.982 & 856.0 & 0.974 \\
        & 1000 & 971.2 & 1171.7 & 954.0 & 0.982 & 946.0 & 0.974 \\
        & 1100 & 1062.6 & 1276.2 & 1044.6 & 0.983 & 1037.8 & 0.977 \\
        \hline 
    \end{tabular}
    \captionsetup{justification=raggedright,singlelinecheck=false}
    \caption{\textbf{Observed performance with respect to heuristic solvers}. Simulations of {our approach} are used to approximate solutions for Max-3SAT instances with 70, 90, or 110 variables and a given number of clauses. For each problem size, results are compared to winning heuristic solutions from the MaxSAT evaluation \cite{maxsat_2016} and averaged over five problem instances. Ratios are given with respect to the classical heuristic solution. \label{tab:heuristic_sim}}
\end{table*}

While these initial simulations of {our approach} for Max-3SAT yield objectives that are slightly less than those given by existing classical heuristic solvers, the advantage of our approach lies in its theoretical basis and generalizability. Classical incomplete solvers may be specifically tailored to MaxSAT, but the PSL framework can be generalized to any polynomial optimization problem.

\section{Our Approach to Max-\textit{k}SAT}

Max-$k$SAT can be expressed as a degree-2$D$ polynomial optimization problem, where $2D = 2\,\text{ceil}(k/2)$ is the maximum degree of the polynomial objective.
The polynomial consists of only even-degree terms of degree-$2d$, where $d \in \{1, ..., D\}$.
The generalized problem is written as
\begin{mini}
    {\rho}{\sum_{d = 1}^D \expval{W^{(d)}, \rho^{\otimes d}}}
    {}{\label{eqn:maxksat_sdp}}
    \addConstraint{\rho_{i,i} = 2^{-n},  \quad \forall i \leq N}
\end{mini}
where the objective function is a sum of expectation values
\begin{align}
    \sum_{d = 1}^D \expval{W^{(d)}, \rho^{\otimes d}} = \sum_{d = 1}^D \bra{\Psi^{(d)}} W^{(d)}  \ket{\Psi^{(d)}}
\end{align}
and we define $\ket{\Psi^{(d)}} \equiv \ket{\psi}^{\otimes d}$ to simplify notation. 

Each $\bra{\Psi^{(d)}} W^{(d)} \ket{\Psi^{(d)}}$ encodes terms of degree-$2d$ and can be estimated using the Hadamard test, 
\begin{align}
    \alpha\bra{\Psi^{(d)}} W^{(d)}  \ket{\Psi^{(d)}} \approx \text{Im}(\bra{\Psi^{(d)}}U_{W^{(d)}}\ket{\Psi^{(d)}})
\end{align}
where $U_{W^{(d)}} \equiv e^{i \alpha W^{(d)}}$ (see Figure~\ref{fig:hadamard_test}). 
We evaluate all terms of degree-$2d$ by simultaneously and separately preparing $d$ identical copies of $\rho$ {and using} the $nd$-qubit unitary $U_{W^{(d)}}$ to conduct the Hadamard test upon the entire quantum system. 
{By} repeating this for all possible values of $d \in \{1, ..., D\}$, we can evaluate all terms in the objective function. 
We therefore require $O(D) = O(k)$ Hadamard tests and $O(nD) = O(nk)$ qubits to estimate the objective function.
This allows us to generalize the original framework~\cite{Patti_2023} for degree-$k$ polynomial optimization with $O(k)$ time and resource complexity.

The Pauli strings and the population-balancing unitary are formulated identically to those in Max-3SAT, since the equality constraint in the optimization problem is the same.

The unrounded SE solution can be generalized to
\begin{align}
    \frac{2^k - 1}{2^k} n_C - \sum_{d = 1}^{D} \frac{2^{nd-1}}{\alpha}\expval{\sigma_\text{anc.}^z}_{W^{(d)} },
\end{align}
where each term may be evaluated with a Hadamard test. 
We require $O(k)$ Hadamard tests to determine this unrounded solution.

The second rounding method ST follows the same procedure described as {in Max-2SAT and Max-3SAT}.

\section{Discussion and Future Work}

We present product-state lifting (PSL) as a simple way to upgrade variational quantum semidefinite programs (vQSDPs) using basis-state encodings to address higher-degree polynomial objectives with linear constraints.
Our approach requires only a linear increase in resources, while keeping the variable mapping and constraint machinery exactly as in the quadratic case.

In this manuscript, we explicitly show how PSL is used with Ref.~\cite{Patti_2023} to tackle degree-$k$ optimizations. 
Beyond Ref.~\cite{Patti_2023}, PSL is generic, and any basis-encoding vQSDP whose loss is written as expectations of Hermitian inputs with linear-constraint penalties can be lifted by swapping objective terms for their higher-order degree-2$d$ counterparts acting on $\rho^{\otimes d}$.
This applies directly to Patel-Coles-Wilde's augmented-Lagrangian formulations (ECSF and ICSF) and their VQAs over purified states, which already optimize unconstrained expectation-values losses~\cite{Patel_Coles_Wilde_2021}.
PSL can also be applied to QSlack, which converts (in)equality constraints with slack variables and penalties and even supports a primal-dual \say{sandwich}~\cite{Chen_Westerheim_Holmes_Luo_Nuradha_Patel_Rethinasamy_Wang_Wilde_2025}.
In both cases, PSL only modifies the objective observables.
The constraint map on the base state and penalty scaffolding are unchanged, suggesting that algorithmic and convergence-style guarantees in those frameworks carry over (with rescaling modifications).
However, formal certificates specific to SDPs, such as QSlack's bounds for SDP optima, are not inherited automatically and we leave any potential PSL analogue to future work.

\vspace{5mm}

{While our current experiments are intended as proof-of-principle demonstrations of PSL within a vQSDP pipeline, a natural direction for future work is a broader empirical study on newer and harder benchmark suites (e.g., recent MaxSAT evaluations
~\cite{maxsat_eval}) together with updated state-of-the-art classical baselines.}
The generalization to any polynomial optimization may {also} be tested by applying {our} approach to other families of polynomial optimization problems, including those that are optimized over \emph{continuous} variables. In these continuous cases, the rounding step of the ST solution method may not be required. One possible application is in quantum chemistry, where the SOS hierarchy (i.e., the reduced density matrix method) is used to study weak coupling perturbation theory \cite{Hastings_2024}. SOS is also used as an approach towards the best separable quantum state problem \cite{Doherty_Parrilo_Spedalieri_2004}.

\vspace{5mm}

Additionally, the high performance of the ST method in {our proof-of-principle} classical simulations suggests that Patti et al.'s algorithm~\cite{Patti_2023} with PSL or similar PSL-inspired methods may be strong candidates for a quantum-inspired algorithm. 
Unrounded SE solutions are not always informative, as they can exceed the total number of clauses as they do in our simulations for Max-3SAT.
In practice, rounding is easier to perform classically rather than through tomography.
Furthermore, the circuit does not need to be implemented on qubits so the Hadamard test is not required, and we may directly evaluate the objective function using the weight matrix.
The approximations induced by converting the weight matrix to a unitary and encoding variables into an exponentially-sized state vector would also no longer be required.
Resulting rounded states could be used as warm-starts for other solvers.
QSlack's companion CSlack further underscores this variational-penalty template on the classical side~\cite{Chen_Westerheim_Holmes_Luo_Nuradha_Patel_Rethinasamy_Wang_Wilde_2025}.

In summary, the PSL formulation is highly generalizable and shows promise for efficiently tackling higher-degree polynomial optimization challenges.
Three immediate directions of future work are (i) formalizing carry-over guarantees, such as PSL analogues of QSlack's bounds or Patel et al.'s first-order convergence statements under our new lifted observables, (ii) testing higher-degree polynomials (including those associated with MaxSAT or other problems) and larger instances with stronger classical outer optimizers, including simulated-annealing-style schedules in place of simple gradient descent, and (iii) formulating and benchmarking quantum-inspired PSL variants.

\textbf{Acknowledgements}\\
R.B. was supported by NSF CCF (grant \#1918549). S.F.Y. would like to thank the NSF via QIdeas HDR (OAC-2118310) and the CUA PFC (PHY-2317134). 
\bibliography{sources}

\bibliographystyle{quantum}

\onecolumn\newpage

\appendix

\section{Sum-of-Squares (SOS) and Rounding Procedures} \label{app:SOS}
In this section of the appendix, we provide more details on the sum-of-squares (SOS) method as well as its rounding procedures.

\subsection{SOS Hierarchy} \label{app:sos_hierarchy}

While there exist several relaxation algorithms and rounding procedures used to approximate solutions to Max-$k$SAT and similar problems, we choose SOS as the most standard and systematic method as a classical benchmark for our algorithm.

Let $F(y)$ be a multivariate polynomial, where $y$ denotes the $N$ variables $\{y_i\}_{i \in \{0, 1, ..., N-1\}}$. The goal of classical SOS is to write $F(y)$ in such a way that the nonnegativity of $F(y)$ becomes obvious. Specifically, the existence of an SOS decomposition 
\begin{align}
    F(y) = \sum_j q_j^2(y),
\end{align}
where $q_j(y)$ are some polynomials in $y$, is an algebraic certificate of nonnegativity.

The SOS hierarchy is a hierarchy of convex relaxations, ordered with increasing power and computational cost. The $D$th level of the hierarchy corresponds to polynomial optimization problems with a degree-$2D$ polynomial objective. The first level of the hierarchy, where $D = 1$, corresponds to an SDP for a quadratic optimization problem. To consider higher-degree polynomials, new variables and constraints must be introduced such that the problem may then be solved as a larger SDP with additional constraints. Specifically, we draw a connection between SOS and semidefinite matrices with the following theorem.

\begin{thm}
    A degree-2$D$ multivariate polynomial $F(y)$ is a sum-of-squares if and only if there exists positive semidefinite matrix $Q$ such that
    \begin{align}
        F(y) = b^T Q b
    \end{align}
    where $b = \begin{pmatrix} 1 & y_1 & y_2 & ... & y_N & y_1 y_2 & ...& y_N^D\end{pmatrix}^T$. The vector $b$ is the polynomial basis, and the matrix $Q$ is called the Gram matrix.
    \label{thm:one}
\end{thm}

A proof is provided by \cite{Princeton_SOS}.

\vspace{5mm}

To show how SOS is applied to polynomial optimization problems, we consider a generic problem of the form
\begin{maxi}
    {}{p(y) \hspace{3.5cm}}
    {}{\label{eqn:sdp_max2SAT}}
    \addConstraint{h_j(y) = 0, \quad \forall j \in \{1, 2, ..., M\}}
\end{maxi}
where $p(y)$ is a polynomial objective that is not necessarily convex, $h_j(y) = 0$ are equality constraints, and $M$ is the number of constraints. Using the given problem, we define a new polynomial\footnote{SOS is often used to determine a lower bound on the solution of the original problem. Here, we want to find an upper bound. To make the notation more straightforward, we introduced a factor of $-1$ to $p(y)$ in our formulation of $F(y)$.}
\begin{align}
    F(y) \equiv - p(y) - \gamma + \sum_j t_j(y) h_j(y),
    \label{eqn:sos_F(y)}
\end{align}
where $\gamma$ is a real-valued variable, and $t_j(y)$ is a polynomial with its degree limited by $D$. If $F(y)$ may be factored into a sum of squares for some value of $\gamma$ and some polynomials $t_j(y)$, we guarantee that $F(y) \geq 0$. When the constraints are satisfied such that $h_j(y) = 0$, we obtain $p(y) \leq - \gamma$ such that $-\gamma$ forms an upper bound on $p(y)$. Formally, we want to obtain a tight upper bound on $p(y)$ by solving
\begin{maxi}
    {}{\gamma \hspace{2cm}}
    {}{\label{eqn:SOS}}
    \addConstraint{F(y) \text{ is SOS.}}
\end{maxi}
Since we look for solutions where $h_j(y) = 0$, the polynomials $t_j(y)$ in Eq.~\eqref{eqn:sos_F(y)} function as additional degrees of freedom that allow tighter upper bounds.

\vspace{5mm}

As a very simple example, let us consider a Max-3SAT problem with two clauses: $(x_1 \lor x_2 \lor x_3)$ and $(\neg x_1 \lor \neg x_2 \lor x_3)$. The equivalent polynomial optimization problem is
\begin{maxi}
    {}{\frac{1}{4}\left(y_0y_3 - y_1 y_2 + y_0y_1y_2y_3 \right) + \frac{7}{4}}
    {}{}
    \addConstraint{y_i^2 = 1\, \forall i \in \{0, 1, 2, 3\}}
\end{maxi}
where the objective function is formulated using Eq.~\eqref{eqn:max3sat_v}. The result may be approximated using the SOS in Eq.~\eqref{eqn:SOS}, where
\begin{align}
\begin{split}
    F(y) \equiv &-\frac{1}{4}\left(y_0y_3 - y_1 y_2 + y_0y_1y_2y_3 \right) - \frac{7}{4} \\
    & - \gamma + \sum_{i = 0}^3 t_i(y) (y_i^2 - 1).
\end{split}
\end{align}
This results in a polynomially-sized SDP by Theorem~\ref{thm:one}. Specifically, we define
\begin{align}
    b \equiv \begin{pmatrix} 1 & y_0 & y_1 & y_2 & y_3 & y_0y_1 & y_2y_3 \end{pmatrix}^T
\end{align}
and determine $Q$ as a function of $\gamma$ and the coefficients of $t_i(y)$ such that $F(y)$ is SOS, that is, $F(y) = b^T Q b = \expval{bb^T, Q^T}$. In this way, the problem becomes the SDP
\begin{maxi}
    {Q \in \mathbb{S}^+}{\expval{W, Q} \hspace{2cm}}
    {}{}
    \addConstraint{\expval{bb^T, Q} = F(y)}
\end{maxi}
where $W$ is defined such that $\expval{W, Q} \propto \gamma$. Solving the SDP, we obtain an upper bound $-\gamma = 2$, as we expect.

\subsection{Rounding Procedures}  \label{app:rounding_procedures}
The rounding procedure used in our SOS solution is equivalent to one of the rounding procedures described by van Maaren et al. \cite{vanMaaren_vanNorden_Heule_2008}, and the steps are as follows. It can be shown that the optimal solution $Q$ of the SOS method described in the previous section has an eigenvalue zero. We determine the orthogonal basis of eigenvectors $v_1, ..., v_L$ of $Q$ that have eigenvalue zero, such that $v_l^T Q v_l = 0$ for $l \in \{1, ..., L\}$. Then, we randomly sample a vector $\lambda = \begin{pmatrix} \lambda_1 & ... & \lambda_L \end{pmatrix}$ from the $L$-dimensional unit sphere and use this to generate a linear combination of these eigenvectors,
\begin{align}
    P = \sum_{l = 1}^l \lambda_l v_l.
\end{align}
The elements of $P$ are rounded to $+1$ or $-1$, and the results are used as the solution $\{y_i\}_{i \in \{0, 1, ... , N-1\}}$ to the original polynomial optimization problem.

We note that the vectors $v_l$ for $l \in \{1, ..., L\}$ do not preserve algebraic consistency, that is, the product of elements corresponding to $y_i$ and $y_j$ do not necessarily equal the element corresponding to $y_i y_j$. There are methods to mitigate this, such as those developed by Ref. \cite{vanMaaren_vanNorden_Heule_2008}. In their experiments using Max-3SAT problems with 20 variables, the rounded result reaches 97.2\% of the upper bound.

\section{Simulation Details}

\begin{figure*}
    \centering
    \includegraphics[width=\textwidth]{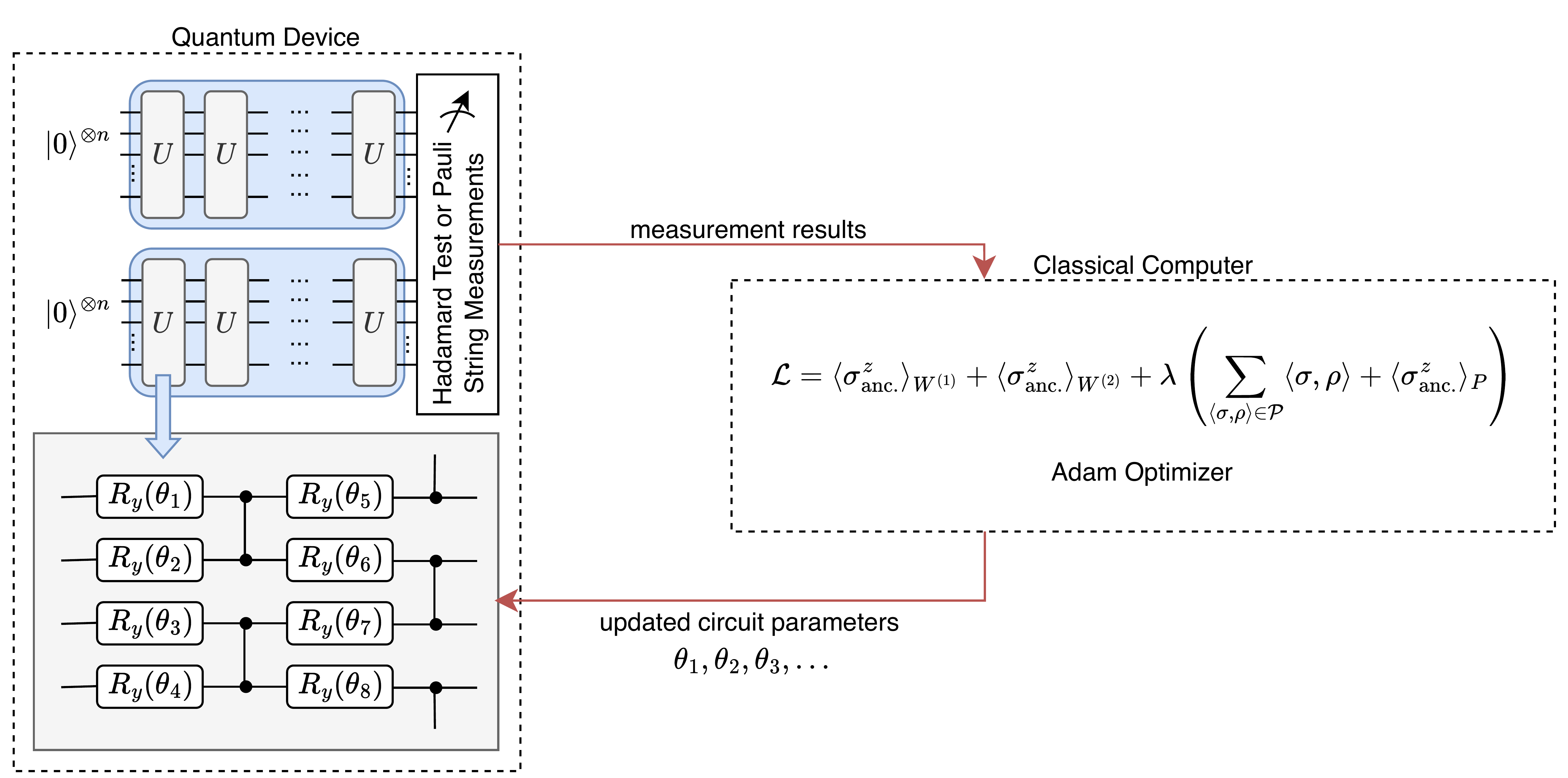}
    \captionsetup{justification=raggedright,singlelinecheck=false}
    \caption{{\textbf{Schematic of the simulated algorithm.} 
    The quantum device (left) holds two identical $n$-qubit quantum registers, on which several layers of the parameterized circuit $U$ are performed. Hadamard tests and Pauli string measurements are performed on the final state, and results are sent to the classical computer (right). The classical computer calculates loss and performs gradient descent to return updated circuit parameters back to the quantum device.}}
    \label{fig:sim_details}
\end{figure*}

{The simulated algorithm for Max-3SAT is shown in Fig.~\ref{fig:sim_details}. Our approach is a variational quantum algorithm, which is a hybrid approach with both a quantum device and classical computer. The quantum device produces and measures a (randomly initialized) parameterized quantum state. Results are sent to the classical computer, which produces the loss function and uses an Adam optimizer to minimize the loss. The cycle of producing the parameterized quantum state, sending measurement results to the classical computer, evaluating loss and performing gradient descent, and sending updated parameters back to the quantum device to use in the next round is repeated over several epochs.}

{The quantum device hosts the quantum state ansatz, prepared using a parameterized circuit. The parameterized circuit we use consists of several layers, where each layer $U$ consists of (1) single qubit rotations $R_y(\theta)$ on each qubit about the $y$-axis by angle $\theta$, (2) CZ gates on even pairs of qubits, (3) another round of $R_y(\theta)$ rotations, and (4) CZ gates on odd pairs of qubits. Each $R_y(\theta)$ corresponds to an independent parameter $\theta$. The quantum state is measured, and results are fed into a classical computer to evaluate a loss function. The classical computer uses an Adam optimizer to minimize loss.}

{The quantum state and circuit are simulated with TensorLy-Quantum~\cite{patti2021tensorlyquantumquantummachinelearning}, and optimization is done with PyTorch.}

{More detailed simulation results are given in Tables~\ref{tab:heuristic_sim_long_v20},~\ref{tab:heuristic_sim_long_v70},~\ref{tab:heuristic_sim_long_v90} and~\ref{tab:heuristic_sim_long_v110}.}

{\begin{table*}
    \centering
    \begin{tabular}{|c||c|c||c||c|c|c|c|}
        \hline
        \multicolumn{1}{|c||}{Problem} & \multicolumn{2}{c||}{Classical SOS} & \multicolumn{1}{c||}{Sim Unround} & \multicolumn{4}{c|}{Sim Round}\\
        \hline \hline
        \# clauses & Unround & Round & Avg Soln & Best Soln & Best Ratio & Avg Soln & Avg Ratio \\
        \hline
        \midrule
  v20c80-1 & 79.1 &      79 &     82.4 &   79.0 &  1.000 &   77.0 &  0.975 \\
  v20c80-2 & 80.0 &      80 &     81.6 &   79.0 &  0.988 &   76.6 &  0.958 \\
  v20c80-3 & 80.0 &      80 &     83.5 &   80.0 &  1.000 &   78.2 &  0.977 \\
  v20c80-4 & 80.0 &      80 &     81.9 &   79.0 &  0.988 &   77.6 &  0.971 \\
  v20c80-5 & 80.1 &      80 &     86.2 &   80.0 &  1.000 &   79.0 &  0.988 \\
  \hline
 v20c100-1 &               100.0 &      89 &    102.1 &   98.0 &  1.101 &   96.4 &  1.083 \\
 v20c100-2 & 99.0 &      99 &    101.3 &   98.0 &  0.990 &   96.5 &  0.975 \\
 v20c100-3 &               100.0 &     100 &    101.0 &   99.0 &  0.990 &   96.2 &  0.962 \\
 v20c100-4 & 99.0 &      87 &    103.0 &   99.0 &  1.138 &   96.1 &  1.105 \\
 v20c100-5 &               100.0 &      88 &    100.9 &  100.0 &  1.136 &   96.0 &  1.091 \\
 \hline
 v20c120-1 &               117.0 &     117 &    118.8 &  117.0 &  1.000 &  114.1 &  0.975 \\
 v20c120-2 &               117.0 &     117 &    119.6 &  117.0 &  1.000 &  114.5 &  0.979 \\
 v20c120-3 &               119.0 &     119 &    125.2 &  119.0 &  1.000 &  117.0 &  0.983 \\
 v20c120-4 &               118.0 &     118 &    118.0 &  117.0 &  0.992 &  113.8 &  0.964 \\
 v20c120-5 &               119.0 &      93 &    123.6 &  119.0 &  1.280 &  115.6 &  1.244 \\
 \hline
 v20c140-1 &               139.0 &     139 &    145.4 &  138.0 &  0.993 &  135.6 &  0.976 \\
 v20c140-2 &               136.0 &     136 &    139.1 &  136.0 &  1.000 &  133.0 &  0.978 \\
 v20c140-3 &               138.0 &     124 &    138.5 &  137.0 &  1.105 &  132.4 &  1.068 \\
 v20c140-4 &               138.0 &     121 &    141.9 &  138.0 &  1.140 &  134.2 &  1.109 \\
 v20c140-5 &               138.0 &     116 &    140.7 &  136.0 &  1.172 &  134.0 &  1.155 \\
 \hline
 v20c160-1 &               157.0 &     156 &    161.6 &  156.0 &  1.000 &  153.3 &  0.983 \\
 v20c160-2 &               156.0 &     155 &    159.3 &  156.0 &  1.006 &  152.4 &  0.983 \\
 v20c160-3 &               157.0 &     156 &    158.0 &  155.0 &  0.994 &  151.6 &  0.971 \\
 v20c160-4 &               157.0 &     156 &    158.4 &  153.0 &  0.981 &  150.0 &  0.962 \\
 v20c160-5 &               155.0 &     142 &    162.8 &  156.0 &  1.099 &  153.4 &  1.080 \\
 \hline
 v20c180-1 &               176.0 &     175 &    183.0 &  175.0 &  1.000 &  171.3 &  0.979 \\
 v20c180-2 &               173.0 &     160 &    182.1 &  178.0 &  1.112 &  173.9 &  1.087 \\
 v20c180-3 &               172.0 &     165 &    177.5 &  172.0 &  1.042 &  168.6 &  1.022 \\
 v20c180-4 &               175.0 &     166 &    174.8 &  171.0 &  1.030 &  168.4 &  1.014 \\
 v20c180-5 &               173.0 &     172 &    177.3 &  172.0 &  1.000 &  169.1 &  0.983 \\
 \hline
    \end{tabular}
    \caption{{\textbf{Detailed simulation results of Max-3SAT with 20 variables}. Problem instance names are given by v[\textit{\# vars}]c[\textit{\# clauses}]-[\textit{instance \#}]. Each row corresponds to one problem instance, and simulation results are averaged over 20 repetitions with different random initializations.} {Ratios are given with respect to the rounded classical SOS solutions.} \label{tab:heuristic_sim_long_v20}}
\end{table*}}

{\begin{table*}
    \centering
    \begin{tabular}{|c||c||c||c|c|c|c|}
        \hline
        \multicolumn{1}{|c||}{Problem} & \multicolumn{1}{c||}{Classical} & \multicolumn{1}{c||}{Sim Unround} & \multicolumn{4}{c|}{Sim Round}\\
        \hline \hline
        Name & Heuristic & Avg Soln & Best Soln & Best Ratio & Avg Soln & Avg Ratio \\
        \hline 
        s3v70c700-1 & 679 & 819.9 & 671 & 0.988 & 665.6 & 0.980  \\
        s3v70c700-2 & 679 & 823.4 & 674 & 0.993 & 664.0 & 0.978 \\
        s3v70c700-3 & 675 & 816.7 & 668 & 0.990 & 660.0 & 0.978 \\
        s3v70c700-4 & 676 & 811.1 & 670 & 0.991 & 666.5 & 0.986 \\
        s3v70c700-5 & 678 & 804.3 & 666 & 0.982 & 661.4 & 0.975 \\
        \hline
        s3v70c800-1 & 769 & 903.1 & 760 & 0.988 & 755.6 & 0.983 \\
        s3v70c800-2 & 766 & 885.2 & 757 & 0.988 & 749.3 & 0.978 \\
        s3v70c800-3 & 770 & 901.3 & 757 &  0.983 & 751.2 & 0.976 \\
        s3v70c800-4 & 772 & 904.9 & 764 & 0.990 & 755.4 & 0.979 \\
        s3v70c800-5 & 769 & 899.8 & 757 & 0.984 & 751.9 & 0.978 \\
        \hline
        s3v70c900-1 & 861 & 1004.0 & 853 & 0.991 & 846.2 & 0.983 \\
        s3v70c900-2 & 862 & 1004.5 & 855 & 0.992 & 849.0 & 0.985 \\
        s3v70c900-3 & 861 & 985.6 & 851 & 0.988 & 839.8 & 0.975 \\
        s3v70c900-4 & 861 & 1002.1 & 851 & 0.988 & 843.5 & 0.980 \\
        s3v70c900-5 & 860 & 1004.9 & 853  & 0.992 & 846.8 & 0.985 \\
        \hline
        s3v70c1000-1 & 966 & 1110.1 & 940 & 0.973 & 935.7 & 0.969 \\
        s3v70c1000-2 & 962 & 1127.7 & 941 & 0.978 & 934.5 & 0.971 \\
        s3v70c1000-3 & 961 & 1105.5 & 939 & 0.977 & 934.2 & 0.972 \\
        s3v70c1000-4 & 966 & 1091.9 & 939 & 0.972 & 932.7 & 0.966 \\
        s3v70c1000-5 & 959 & 1138.4 & 944 & 0.984 & 938.2 & 0.978 \\
        \hline
        s3v70c1100-1 & 1044 & 1206.2 & 1034 & 0.990 & 1028.2 & 0.985 \\
        s3v70c1100-2 & 1045 & 1226.5 & 1040 & 0.995 & 1031.4 & 0.987 \\
        s3v70c1100-3 & 1047 & 1211.6 & 1039 & 0.992 & 1031.6 & 0.985 \\
        s3v70c1100-4 & 1048 & 1219.8 & 1042 & 0.994 & 1035.0 & 0.988 \\
        s3v70c1100-5 & 1047 & 1208.0 & 1041 & 0.994 & 1029.5 & 0.983 \\
        \hline
        s3v70c1200-1 & 1134 & 1304.0 & 1125 & 0.992 & 1116.8 & 0.985 \\
        s3v70c1200-2 & 1137 & 1329.2 & 1126 & 0.990 & 1120.8 & 0.986 \\
        s3v70c1200-3 & 1135 & 1312.4 & 1126 & 0.992 & 1117.2 & 0.984 \\
        s3v70c1200-4 & 1133 & 1303.7 & 1122 & 0.990 & 1112.7 & 0.982 \\
        s3v70c1200-5 & 1135 & 1340.5 & 1121 & 0.988 & 1116.4 & 0.984 \\
        \hline
        s3v70c1300-1 & 1225 & 1382.6 & 1213 & 0.990 & 1206.0 & 0.984 \\
        s3v70c1300-2 & 1226 & 1423.4 & 1212 & 0.989 & 1207.2 & 0.985 \\
        s3v70c1300-3 & 1230 & 1402.7 & 1215 & 0.988 & 1208.2 & 0.982 \\
        s3v70c1300-4 & 1232 & 1401.9 & 1216 & 0.987 & 1208.3 & 0.981 \\
        s3v70c1300-5 & 1230 & 1477.0 & 1220 & 0.992 & 1212.8 & 0.986 \\
        \hline
        s3v70c1400-1 & 1322 & 1515.6 & 1305 & 0.987 & 1297.2 & 0.981 \\
        s3v70c1400-2 & 1317 & 1517.6 & 1307 & 0.992 & 1296.3 & 0.984 \\
        s3v70c1400-3 & 1321 & 1512.5 & 1313 & 0.994 & 1297.6 & 0.982 \\
        s3v70c1400-4 & 1325 & 1523.6 & 1309 & 0.988 & 1302.2 & 0.983 \\
        s3v70c1400-5 & 1318 & 1538.1 & 1306 & 0.991 & 1301.0 & 0.987 \\
        \hline
        s3v70c1500-1 & 1410 & 1632.2 & 1394 & 0.989 & 1386.9 & 0.984 \\
        s3v70c1500-2 & 1411 & 1617.1 & 1397 & 0.990 & 1390.8 & 0.986 \\
        s3v70c1500-3 & 1414 & 1617.4 & 1401 & 0.991 & 1393.0 & 0.985 \\
        s3v70c1500-4 & 1409 & 1608.4 & 1399 & 0.993 & 1390.1 & 0.987 \\
        s3v70c1500-5 & 1406 & 1595.2 & 1396 & 0.993 & 1385.2 & 0.985 \\
        \hline 
    \end{tabular}
    \caption{{\textbf{Detailed simulation results of Max-3SAT with 70 variables}. Problem instance names are given by s[\textit{\# vars per clause}]v[\textit{\# vars}]c[\textit{\# clauses}]-[\textit{instance \#}]. Each row corresponds to one problem instance, and simulation results are averaged over 20 repetitions with different random initializations. Classical heuristic results are the best solutions from~\cite{maxsat_2016}.} {Ratios are given with respect to the classical heuristic solutions.} \label{tab:heuristic_sim_long_v70}}
\end{table*}}

\begin{table*}
    \centering
    \begin{tabular}{|c||c||c||c|c|c|c|}
        \hline
        \multicolumn{1}{|c||}{Problem} & \multicolumn{1}{c||}{Classical} & \multicolumn{1}{c||}{Sim Unround} & \multicolumn{4}{c|}{Sim Round}\\
        \hline \hline
        Name & Heuristic & Avg Soln & Best Soln & Best Ratio & Avg Soln & Avg Ratio \\
        \hline
        s3v90c700-1 & 684 & 831.5 & 676 & 0.988 & 666.1 & 0.974 \\
        s3v90c700-2 & 683 & 825.3 & 670 & 0.981 & 663.0 & 0.971 \\
        s3v90c700-3 & 685 & 840.0 & 674 & 0.984 & 668.5 & 0.976 \\
        s3v90c700-4 & 682 & 809.9 & 670 & 0.982 & 666.2 & 0.977 \\
        s3v90c700-5 & 684 & 823.0 & 669 & 0.978 & 663.6 & 0.970 \\
        \hline
        s3v90c800-1 & 778 & 936.3 & 760 & 0.977 & 755.0 & 0.970 \\
        s3v90c800-2 & 777 & 966.5 & 765 & 0.985 & 758.6 & 0.976 \\
        s3v90c800-3 & 777 & 958.5 & 768 & 0.988 & 762.8 & 0.982 \\
        s3v90c800-4 & 780 & 931.8 & 767 & 0.983 & 761.6 & 0.976 \\
        s3v90c800-5  & 776 & 936.8 & 765 & 0.986 & 757.8 & 0.977 \\
        \hline
        s3v90c900-1 & 872 & 1038.9 & 856 & 0.982 & 850.0 & 0.975 \\
        s3v90c900-2 & 872 & 1028.3 & 859 & 0.985 & 850.6 & 0.975 \\
        s3v90c900-3 & 868 & 1019.5 & 856 & 0.986 & 849.3 & 0.978 \\
        s3v90c900-4 & 872 & 1052.6 & 861 & 0.987 & 852.6 & 0.978 \\
        s3v90c900-5 & 875 & 1048.2 & 858 & 0.981 & 851.3 & 0.973 \\
        \hline
        s3v90c1000-1 & 966 & 1148.8 & 956 & 0.990 & 942.6 & 0.976 \\
        s3v90c1000-2 & 962 & 1166.9 & 946 & 0.983 & 940.6 & 0.978 \\
        s3v90c1000-3 & 961 & 1129.7 & 943 & 0.981 & 936.2 & 0.974 \\
        s3v90c1000-4 & 966 & 1149.4 & 953 & 0.987 & 944.9 & 0.978 \\
        s3v90c1000-5 & 959 & 1138.4 & 947 & 0.987 & 940.6 & 0.981 \\
        \hline
        s3v90c1100-1 & 1056 & 1235.5 & 1041 & 0.986 & 1032.0 & 0.977 \\
        s3v90c1100-2 & 1056 & 1245.8 & 1040 & 0.985 & 1032.3 & 0.978 \\
        s3v90c1100-3 & 1053 & 1235.8 & 1042 & 0.990 & 1030.6 & 0.979 \\
        s3v90c1100-4 & 1057 & 1247.2 & 1044 & 0.988 & 1037.2 & 0.981 \\
        s3v90c1100-5 & 1056 & 1238.8 & 1040 & 0.985 & 1031.6 & 0.977 \\
        \hline
        s3v90c1200-1 & 1153 & 1401.0 & 1138 & 0.987 & 1128.1 & 0.978 \\
        s3v90c1200-2 & 1146 & 1340.9 & 1132 & 0.988 & 1122.4 & 0.979 \\
        s3v90c1200-3 & 1150 & 1354.4 & 1140 & 0.991 & 1129.4 & 0.982 \\
        s3v90c1200-4 & 1145 & 1339.8 & 1134 & 0.990 & 1124.4 & 0.982 \\
        s3v90c1200-5 & 1141 & 1333.3 & 1129 & 0.989 & 1116.0 & 0.978 \\
        \hline
        s3v90c1300-1 & 1240 & 1442.8 & 1219 & 0.983 & 1211.2 & 0.977 \\
        s3v90c1300-2 & 1236 & 1430.4 & 1221 & 0.988 & 1214.0 & 0.982 \\
        s3v90c1300-3 & 1236 & 1442.8 & 1219 & 0.986 & 1212.7 & 0.981 \\
        s3v90c1300-4 & 1237 & 1436.4 & 1216 & 0.983 & 1210.0 & 0.978 \\
        s3v90c1300-5 & 1242 & 1440.5 & 1229 & 0.990 & 1219.0 & 0.981 \\

        \hline
    \end{tabular}
    \caption{{\textbf{Detailed simulation results of Max-3SAT with 90 variables}. Problem instance names are given by s[\textit{\# vars per clause}]v[\textit{\# vars}]c[\textit{\# clauses}]-[\textit{instance \#}]. Each row corresponds to one problem instance, and simulation results are averaged over 20 repetitions with different random initializations. Classical heuristic results are the best solutions from~\cite{maxsat_2016}.} {Ratios are given with respect to the classical heuristic solutions.}  \label{tab:heuristic_sim_long_v90}}
\end{table*}

{\begin{table*}
    \centering
    \begin{tabular}{|c||c||c||c|c|c|c|}
        \hline
        \multicolumn{1}{|c||}{Problem} & \multicolumn{1}{c||}{Classical} & \multicolumn{1}{c||}{Sim Unround} & \multicolumn{4}{c|}{Sim Round}\\
        \hline \hline
        Name & Heuristic & Avg Soln & Best Soln & Best Ratio & Avg Soln & Avg Ratio \\
        \hline
        s3v110c700-1 & 689 & 836.8 & 675 & 0.980 & 668.6 & 0.970 \\
        s3v110c700-2 & 688 & 840.0 & 675 & 0.981 & 668.0 & 0.971 \\
        s3v110c700-3 & 690 & 863.9 & 673 & 0.975 & 670.0 & 0.971 \\
        s3v110c700-4 & 689 & 840.5 & 673 & 0.977 & 667.9 &0.969 \\
        s3v110c700-5 & 690 & 838.7 & 675 & 0.978 & 669.0 & 0.970 \\
        s3v110c800-1 & 783 & 951.8 & 764 & 0.976 & 758.2 & 0.968 \\
        s3v110c800-2 & 783 & 934.5 & 769 & 0.982 & 759.7 & 0.970 \\
        s3v110c800-3 & 785 & 991.0 & 774 & 0.986 & 766.5 & 0.976 \\
        s3v110c800-4 & 782 & 955.6 & 768 & 0.982 & 761.6 & 0.974 \\
        s3v110c800-5 & 783 & 952.6 & 771 & 0.985 & 760.9 & 0.972 \\
        s3v110c900-1 & 878 & 1057.2 & 864 & 0.984 & 855.2 & 0.974 \\
        s3v110c900-2 & 878 & 1074.7 & 866 & 0.986 & 856.7 & 0.976 \\
        s3v110c900-3 & 883 & 1089.5 & 863 & 0.977 & 857.8 & 0.971 \\
        s3v110c900-4 & 879 & 1083.6 & 865 & 0.984 & 856.7 & 0.975 \\
        s3v110c900-5 & 878 & 1056.3 & 861 & 0.981 & 853.6 & 0.972 \\
        s3v110c1000-1 & 970 & 1172.2 & 952 & 0.981 & 941.6 & 0.971 \\
        s3v110c1000-2 & 973 & 1176.2 & 959 & 0.986 & 950.4 & 0.977 \\
        s3v110c1000-3 & 972 & 1189.8 & 955 & 0.983 & 950.0 & 0.977 \\
        s3v110c1000-4 & 971 & 1165.2 & 951 & 0.979 & 944.4 & 0.973 \\
        s3v110c1000-5 & 970 & 1154.7 & 953 & 0.982 & 943.6 & 0.973 \\ 
        s3v110c1100-1 & 1066 & 1295.7 & 1051 & 0.986 & 1044.0 & 0.979 \\
        s3v110c1100-2 & 1063 & 1262.4 & 1039 & 0.977 & 1034.6 & 0.973 \\
        s3v110c1100-3 & 1062 & 1272.3 & 1047 & 0.986 & 1038.0 & 0.977 \\
        s3v110c1100-4 & 1061 & 1275.7 & 1045 & 0.985 & 1035.6 & 0.976 \\
        s3v110c1100-5 & 1061 & 1275.1 & 1041 & 0.981 & 1036.9 & 0.977 \\
        \hline
    \end{tabular}
    \caption{{\textbf{Detailed simulation results of Max-3SAT with 110 variables}. Problem instance names are given by s[\textit{\# vars per clause}]v[\textit{\# vars}]c[\textit{\# clauses}]-[\textit{instance \#}]. Each row corresponds to one problem instance, and simulation results are averaged over 20 repetitions with different random initializations. Classical heuristic results are the best solutions from~\cite{maxsat_2016}.} {Ratios are given with respect to the classical heuristic solutions.}  \label{tab:heuristic_sim_long_v110}}
\end{table*}}

The following are the hyperparameter values used:
\begin{itemize}
    \item Number of epochs: 100
    \item Number of parametrized circuit layers $U$: 100
    \item Learning rate: 0.01
    \item Exponential coefficient $\alpha$: 0.01
    \item Lagrange multiplier $\lambda$: 300$\alpha$/(\# Pauli string constraints)
\end{itemize}

\end{document}